\documentclass[]{tMPH2e}

\usepackage{amssymb}

\usepackage{amsmath}

\usepackage[usenames]{color}

\def\etal{\mbox{et al.}}

\newcommand\beq{\begin{equation}}

\newcommand\eeq{\end{equation}}
\newcommand\beqa{\begin{eqnarray}}
\newcommand\eeqa{\end{eqnarray}}
\newcommand{\nn}{\nonumber\\}

\newcommand{\aQ}{B_0}

\newcommand{\aKQ}{B}

\newcommand{\SE}{A}

\newcommand{\aC}{W}

\def\l{\lambda}
\def\e{\eta}

\def\HS{{\text{HS}}}

\begin{document}
\doi{10.1080/0026897YYxxxxxxxx}
 \issn{1362–3028}
\issnp{0026–8976}
\jvol{00}
\jnum{00} \jyear{2011} 

\markboth{S. B. Yuste, A. Santos and M. L\'opez de Haro}{Molecular Physics}

\title{Structure of the square-shoulder fluid}

\author{S. B. Yuste$^{a}$$^{\ast}$\thanks{$^\ast$Email: santos@unex.es
\vspace{6pt}}, A. Santos$^{a}$$^{\dagger}$\thanks{$^\dagger$Corresponding author. Email: andres@unex.es
\vspace{6pt}} and M. L\'opez de Haro$^{b}$$^{\ddagger}$\thanks{$^\ddagger$Email: malopez@servidor.unam.mx
\vspace{6pt}}\\\vspace{6pt}  $^{a}${\em{Departamento de F\'{\i}sica, Universidad de Extremadura, Badajoz 06071, Spain}};
$^{b}${\em{Centro de Investigaci\'{o}n en Energ\'{\i}a, Universidad Nacional Aut\'onoma de M\'exico (U.N.A.M.), Temixco, Morelos 62580, M{e}xico}} }

\maketitle

\begin{abstract}
The structural properties of square-shoulder fluids are derived from the use of  the rational function approximation method. The computation of both the radial distribution function and the static structure factor involves mostly analytical steps, requiring only the numerical solution of a single transcendental equation. The comparison with available simulation data and with numerical solutions of the Percus--Yevick and hypernetted-chain integral equations shows that the present approximation represents an improvement over the Percus--Yevick theory for this system and a reasonable compromise between accuracy and simplicity.\bigskip

\bigskip

\end{abstract}

\section{Introduction}

It is well known that the inclusion of attractive interactions  in the intermolecular potential used to describe a fluid is crucial to obtain a liquid-vapour transition. Perhaps the simplest model accounting for this fact is the square-well (SW) fluid in which the interaction potential is given by
\begin{equation}
\phi_\text{SW} (r)=\left\{
\begin{array}{ll}
\infty , & r<\sigma, \\
-\epsilon, & \sigma<r<\lambda \sigma, \\
0, & r>\lambda \sigma,
\label{SW}
\end{array}
\right.
\end{equation}
where $r$ is the distance, $\sigma$ is the diameter of the hard core, $\epsilon>0$ is the well depth and $(\lambda - 1) \sigma$ is the well width.
The thermodynamic properties of the SW fluid only depend on three dimensionless parameters, namely the packing fraction $\eta\equiv (\pi/6) \rho \sigma^3$ ($\rho$ being the number density),
the reduced temperature $T^*=k_B T/\epsilon$ ($k_B$ and $T$ being the Boltzmann constant and the absolute temperature, respectively) and the width parameter $\lambda$. Due to the combined assets
of relative simplicity and `realistic' features, the SW fluid has been studied thoroughly using both theoretical approaches and simulations
(see, for example,  \cite{SW1,SW2,SW3,SW4,YS94,SW5,LKLLW99,SW6,A00,AS01,SW7,SW7bis,SW8,SW9,LSAS03,SW10,SW11,SW12,SW13,SW14} and references therein).

Another closely related interaction potential is the `square-shoulder' (SS) potential, that has also been
considered in the literature. It reads
\begin{equation}
\phi_\text{SS} (r)=\left\{
\begin{array}{ll}
\infty , & r<\sigma, \\
\epsilon, & \sigma<r<\lambda \sigma, \\
0, & r>\lambda \sigma.
\label{SS}
\end{array}
\right.
\end{equation}
This purely repulsive potential, apparently considered first by Hemmer and Stell 40 years ago \cite{HS70}, has been the subject of recent attention \cite{BF94,YA77,RVMN97,M98,BMAGPSSX09,ZK01,MP03,PK08,FK10,BSB09,SY76,LK93,GGJB99,RS03,S05,S06,S09,ZS09,GSC10,BL10}. On the one hand, it may lead to an isostructural solid-solid transition \cite{BF94}, to a fluid-solid coexisting line with a maximum melting temperature \cite{YA77},  to unusual phase behaviour \cite{RVMN97,M98,BMAGPSSX09} and to a rich variety of (self-organised) ordered structures \cite{ZK01,MP03,PK08,FK10}. On the other hand, it is the simplest model belonging to the class of core-softened potential models for fluids that have been used to study a number of interesting substances such as water \cite{BSB09}, metallic systems \cite{SY76}, colloidal suspensions \cite{LK93} and aqueous solutions of electrolytes \cite{GGJB99}.

It should also be noted that, {as in the SW case,} the thermodynamic properties of the SS system only depend on the packing fraction $\eta$, the reduced temperature $T^*$ and the width parameter $\lambda$. Further, the SS potential becomes equivalent to a hard-sphere (HS) interaction of diameter $\sigma$ in the high-temperature limit {($T^*\to \infty$) or in  the limit of vanishing shoulder width  ($\lambda\to 1$)}, and to an HS interaction of diameter $\l\sigma$ in the low-temperature limit ($T^*\to 0$).

Despite the simplicity of the SS potential, so far no exact results for the thermodynamic or structural properties of this system are available. Further, even in the simplest approximation within the integral equation approach for the study of liquids, namely the Percus--Yevick (PY) closure for the Ornstein--Zernike (OZ) integral equation, no analytical results have been derived for the SS fluid,  Very recently, Zhou and Solana \cite{ZS09} have reported simulations and theoretical results based on a bridge function approximation to close the OZ equation, while Guill\'en-Escamilla {\etal} \cite{GSC10} have also presented simulation results and a parametrization of the direct correlation function which quantitatively agrees with the numerical solution of the OZ equation within the PY approximation.

{Several} years ago two of us \cite{YS94} derived approximate analytical results for the structural properties of the SW fluid using a methodology that has proved useful for many other systems \cite{HYS08}. {Exploiting the fact that this methodology does not make explicit use of the positive character of the well depth, the major aim of this paper is to extend it to the SS  system and compare the resulting structure  with the simulation data available to our knowledge in the literature,} namely those just mentioned by Zhou and Solana \cite{ZS09} and Guill\'en-Escamilla {\etal} \cite{GSC10}, as well as  the data in the paper by Lang {\etal} \cite{LKLLW99}. Also, given the fact that our approach represents an alternative \emph{analytical} route to the integral equation approach for the structural properties of fluids, we will further assess its merits for the SS fluid by comparing our results with those we have obtained from the numerical solution of the OZ equation {both with the PY  and the hypernetted-chain (HNC) closures}.

The paper is organised as follows. In order to make it self-contained, in the next section we present the derivation of the structural properties of the SS fluid using the methodology that {we  refer} to as the Rational Function Approximation (RFA). This is followed in Section {\ref{sec3} by a comparison of our analytical approximation and of the numerical solution of the PY  and HNC equations for the SS potential with the available simulation results. The paper is closed in Section \ref{sec4} with further discussion and concluding remarks.

\section{{Radial distribution function}  of the square-shoulder fluid\label{sec2}}

In this section, we present our proposal for the structural properties of the SS fluid. It follows very closely the parallel derivation for the SW fluid as presented in  \cite{HYS08}.

\subsection{Physical requirements}
As in {previous} works \cite{YS91}, it is convenient to consider the Laplace transform of $rg(r)$, where $g(r)$ is the radial distribution function (rdf), namely
\beq
\label{2.1}
G(s)=\int_0^\infty d r \,e^{-sr}r g(r)
\eeq
and the auxiliary
function $\Psi(s)$ defined through
\beq
\label{2.2}
G(s)=\frac{1}{2\pi}\frac{s}{\rho+e^{s\sigma}\Psi(s)}.
\eeq
The choice of $G(s)$ as the Laplace transform of $r g(r)$ and the
definition of $\Psi(s)$ from Equation \eqref{2.2} are suggested by the
exact form of $g(r)$ to first order in density \cite{YS94}.

Since $g(r)=0$ for $r<\sigma$ while $g(\sigma^+)=\text{finite}$, one
has
\beq
\label{3.2s}
g(r)=\Theta(r-\sigma)\left[g(\sigma^+)+
g'(\sigma^+)(r-\sigma)+\cdots\right],
\eeq
where $g'(r)\equiv d g(r)/d r$ and $\Theta \left(x\right)$ is the Heaviside step function. The foregoing property imposes a constraint
on the large-$s$ behaviour of $G(s)$, namely
\beq
\label{3.3s}
 e^{\sigma s}s G(s)=\sigma g(\sigma^+ )
+\left[g(\sigma^+ )+\sigma g'(\sigma^+)\right] s^{-1}+{\cal
O}(s^{-2}).
\eeq
Therefore, $\lim_{s\to\infty} e^{s\sigma}sG(s)=\sigma
g(\sigma^+)=\text{finite}$ or, equivalently,
\beq
\label{2.3}
\lim_{s\to\infty}s^{-2}\Psi(s)=\frac{1}{2\pi\sigma
g(\sigma^+)}=\text{finite}.
\eeq

On the other hand,  according to the definition of the (reduced) isothermal compressibility
\beqa
\chi &\equiv& k_BT \left(\frac{\partial \rho}{\partial
p}\right)_T\nn
 &=&1+24\eta
\sigma^{-3}\int_0^\infty d r\, r^2 \left[g(r)-1\right],
\label{chi}
\eeqa
it follows that
\beqa
\chi &=&1-24\eta\sigma^{-3}\lim_{s\to 0}\frac{d}{d
s}\int_0^\infty d r\, e^{-s r}r\left[g(r)-1\right]\nn
&=&1-24\eta\sigma^{-3}\lim_{s\to 0}\frac{d}{d
s}\left[G(s)-s^{-2}\right].
\label{GG}
\eeqa
Since the isothermal compressibility $\chi $ is also
finite, one has $\int_0^\infty d r\,r^2\left[g(r)-1\right]=\text{finite}$, so that the \emph{weaker}
condition
 $\int_0^\infty d r \,r\left[g(r)-1\right]=\lim_{s\to 0}[G(s)-s^{-2}]=\text{finite}$
must hold. This in turn implies
\beq
\label{2.4}
\Psi(s)=-\rho+\rho\sigma s-\frac{1}{2}\rho \sigma^2
s^2 +\left(\frac{1}{6}\rho\sigma^3+\frac{1}{2\pi}\right)s^3-
\left(\frac{1}{24}\rho\sigma^3+ \frac{1}{2\pi}\right)\sigma
s^4+{\cal O}(s^5).
\eeq

{Equation
\eqref{2.2} can be formally rewritten as}
\beq
G(s)=-\frac{s}{2\pi}\sum_{n=1}^\infty
\rho^{n-1}\left[-\Psi(s)\right]^{-n} e^{-ns\sigma}.
\eeq
{Thus, the  rdf is given by}
\begin{equation}
g\left({r}\right) =\frac{1}{2\pi r} \sum_{n=1}^{\infty
}\rho^{n-1}\psi _{n}\left(r-n\sigma\right) \Theta
\left(r-n\sigma\right) ,
\label{g(r)}
\end{equation}
with
\begin{equation}
\psi_{n}\left( r\right) =-\mathcal{L}^{-1}\left\{ s\left[ -\Psi
\left( s\right) \right] ^{-n}\right\} ,
\label{varphi}
\end{equation}
{$\mathcal{L}^{-1}$ denoting the inverse Laplace transform.}

{So far, Equations \eqref{2.2}, \eqref{2.3}, \eqref{2.4} and \eqref{g(r)} apply exactly to \emph{any} interaction potential having a hard core at $r=\sigma$. This includes, among other models, the HS, SW and SS potentials.}

For the SS potential given in Equation (\ref{SS}), $G(s)$ must reflect the fact that
$g(r)$ is discontinuous at $r=\lambda \sigma$ as a consequence of the
discontinuity of the potential $\phi_\text{SS}(r)$ and the continuity of the
cavity function $y(r)=\exp\left[{{\phi_\text{SS}(r)}/{k_B T}}\right]g(r)$. This implies that $G(s)$, and hence
$\Psi(s)$, must contain the exponential term $e^{-(\lambda-1)\sigma s}$.
This manifests itself in the low-density limit, where the condition
$\lim_{\rho\to 0}y(r)=1$ yields
\begin{equation}
\lim_{\rho \to 0}g (r)=\left\{
\begin{array}{ll}
0 , & r<\sigma, \\
e^{-1/T^*}, & \sigma<r<\lambda \sigma, \\
1, & r>\lambda \sigma,
\label{SSb}
\end{array}
\right.
\end{equation}
\beq
\lim_{\rho \to 0}G(s)=e^{-1/T^*}e^{-\sigma s}\frac{1+\sigma s}{s^2}+\left(1-e^{-1/T^*}\right)e^{-\lambda\sigma s}\frac{1+\lambda\sigma s}{s^2}.
\label{SSc}
\eeq
{Therefore, from Equation \eqref{2.2} we have}
\beq
\lim_{\rho\to
0}\Psi(s)=\frac{s^3}{2\pi}\left[e^{-1/T^*}(1+s)+e^{-(\lambda-1)s}
(1-e^{-1/T^*})(1+\lambda s )\right]^{-1},
\label{SW1}
\eeq
where we have taken, without loss of generality, $\sigma=1$. This means that in what follows all distances are measured in units of the hard-core diameter $\sigma$.

{Equations \eqref{2.3}, \eqref{2.4} and \eqref{SW1} are basic exact properties that any reasonable approximation for $\Psi(s)$ must fulfill.}

\subsection{The rational function approximation}
In the spirit of the RFA method that has been used in other cases \cite{HYS08}, and following the same rationale as for the SW fluid \cite{YS94},  the simplest form that complies
with Equation \eqref{2.3} and is consistent with Equation \eqref{SW1} is
\begin{equation}
\Psi (s)=\frac{1}{2\pi}\frac{-12\eta+\SE_1 s+\SE_2 s^2+\SE_3
s^3}{1-\aQ+\aKQ_1 s+e^{-(\lambda-1)s}\left(\aQ+\aKQ_2 s\right)},
\label{eq:F(t)}
\end{equation}
where the coefficients $\aQ$, $\aKQ_1$, $\aKQ_2$, $\SE_1$, $\SE_2$
and $\SE_3$ are functions of $\eta$, $T^*$ and $\lambda$.
{Comparison with Equation \eqref{SW1} shows that in the limit $\eta\to 0$ one must have $\SE_1\to 0$, $\SE_2\to 0$, $\SE_3\to 1$, $\aQ\to 1-e^{-1/T^*}$, $\aKQ_1\to e^{-1/T^*}$ and $\aKQ_2\to (1-e^{-1/T^*})\l$.}

The condition
\eqref{2.4} allows one to express the parameters $\aKQ_1$, $\SE_1$,
$\SE_2$ and $\SE_3$ as linear functions of $\aQ$ and $\aKQ_2$ \cite{YS94,AS01}:
\beq
\label{c5}
\aKQ_1=\frac{1}{1+2\eta}\Big[1+\frac{\eta}{2}-2\eta(\lambda^3-1)
\aKQ_2+\frac{\eta}{2}(\lambda-1)^2
(\lambda^2+2\lambda+3) {\aQ}\Big]
 -\aKQ_2+(\lambda-1)\aQ,
\eeq
\beq
\label{c6}
\SE_1=\frac{6\eta^2}{1+2\eta}\Big[{3}+4(\lambda^3-1) \aKQ_2-
(\lambda-1)^2(\lambda^2+2\lambda+3) {\aQ}\Big],
\eeq
\beq
\label{c7}
\SE_2=\frac{6\eta}{1+2\eta}\left\{1-\eta+2(\lambda-1)\left[1-2\eta
\lambda(\lambda+1)\right]\aKQ_2-(\lambda-1)^2\left[(1-\eta(\lambda+1)^2\right]{\aQ}\right\},
\eeq
\beqa
\label{c7bis}
\SE_3&=&\frac{1}{1+2\eta}\Big\{(1-\eta)^2-6\eta(\lambda-1)
\left(\lambda+1-2\eta \lambda^2\right)\aKQ_2\nn && +\eta(
\lambda-1)^2[4+2\lambda-\eta(3\lambda^2+2 \lambda+1)]{\aQ}\Big\}.
\eeqa

{}From Equations (\ref{2.3}) and \eqref{eq:F(t)}, we have
\beq
g(1^+)=\frac{\aKQ_1}{ \SE_3}.
\label{SWB0}
\eeq
{More generally,  the complete rdf follows from Equations \eqref{g(r)}, \eqref{varphi} and \eqref{eq:F(t)}.}
 In particular,
\beq
\psi_1(r)=\psi_{10}(r)\Theta(r)+\psi_{11}(r+1- \lambda)\Theta(r+1- \lambda),
\label{SWB1}
\eeq
\beq
\psi_2(r)=\psi_{20}(r)\Theta(r)+\psi_{21}(r+1- \lambda)\Theta(r+1-
\lambda)+\psi_{22}(r+2-2 \lambda)\Theta(r+2-2 \lambda),
\label{SWB4}
\eeq
where
\beq
\psi_{1k}(r)={2\pi}\sum_{i=1}^3 \frac{\aC_{1k}(s_i)}{\SE'(s_i)}s_i
e^{s_i r},
\label{SWB2}
\eeq
\beq
\psi_{2k}(r)=-4\pi^2\sum_{i=1}^3 \left[r
\aC_{2k}(s_i)+\aC_{2k}'(s_i)-\aC_{2k}(s_i)\frac{\SE''(s_i)}{\SE'(s_i)}\right]\frac{e^{s_i
r}}{[\SE'(s_i)]^2}.
\label{SWB5}
\eeq
Here, $s_i$ are the three  roots of $\SE(s)\equiv
-12\eta+\SE_1 s+\SE_2 s^2+\SE_3s^3$, {the primes denote derivatives with respect to $s$ and the functions $\aC_{1k}(s)$ and  $\aC_{2k}(s)$ are defined as}
\begin{subequations}
\label{SSB}
\beq
\aC_{10}(s)\equiv 1-\aQ+\aKQ_1 s,\quad \aC_{11}(s)\equiv
\aQ+\aKQ_2 s,
\label{SWB3}
\eeq
\beq
\aC_{20}(s)\equiv s[\aC_{10}(s)]^2,\quad \aC_{22}(s)\equiv s[\aC_{11}(s)]^2,
\label{SWB6}
\eeq
\beq
\aC_{21}(s)\equiv 2s
\aC_{10}(s)\aC_{11}(s).
\label{SWB6b}
\eeq
\end{subequations}

To close the proposal, we need to determine the parameters $\aQ$ and
$\aKQ_2$ by imposing two new conditions. An obvious condition is the
continuity of the cavity function at $r=\lambda$, what implies
\beq
g( \lambda^-)=e^{-1/T^*}g( \lambda^+).
\label{SWB8}
\eeq
{From Equations  \eqref{g(r)} and \eqref{SWB1}, assuming $\lambda<2$, one has}
\beq
g(\l^-)=\frac{\psi_{10}(\l-1)}{2\pi\l},
\label{gl-}
\eeq
\beq
g(\l^+)=\frac{\psi_{10}(\l-1)+\psi_{11}(0)}{2\pi\l}.
\label{gl+}
\eeq
{Thus, Equation \eqref{SWB8} yields}
\beq
\psi_{10}(
\lambda-1)=\frac{\psi_{11}(0)}{e^{1/T^*}-1}.
\label{SWB7a}
\eeq
{Equations \eqref{varphi} and \eqref{eq:F(t)} imply $\psi_{11}(0)=2\pi{\aKQ_2}/{ \SE_3}$. Therefore, making use of Equation \eqref{SWB2}, one gets}
\beq
\sum_{i=1}^3 \frac{1-\aQ+\aKQ_1 s_i}{\SE_1 +2\SE_2 s_i+3\SE_3s_i^2}s_i
e^{s_i (\l-1)}=\frac{\aKQ_2}{\left(e^{1/T^*}-1\right) \SE_3}.
\label{SWB7}
\eeq
{Note that, taking into account \eqref{SWB7a}, Equations \eqref{gl-} and \eqref{gl+} can be rewritten as}
\beq
g(\l^-)=\frac{\aKQ_2}{\l\left(e^{1/T^*}-1\right) \SE_3},
\label{gl-b}
\eeq
\beq
g(\l^+)=\frac{\aKQ_2}{\l\left(1-e^{-1/T^*}\right) \SE_3}.
\label{gl+b}
\eeq

As an extra condition, we could enforce the continuity of the first
derivative $y'(r)$ at $r=\lambda$ \cite{A00}. However, this complicates
the problem too much without any relevant gain in accuracy. In
principle, it might be possible to impose consistency with a given
{equation of state},  via either the virial route, the compressibility route, or the
energy route. But this is not practical since no simple {equation of state} for SS
fluids is at our disposal for wide values of density, temperature
and range. As a compromise between simplicity and accuracy, we fix
the parameter $\aQ$ at its exact zero-density limit value, namely \cite{YS94}
\beq
\aQ=1-e^{-1/T^*} .
\label{Q0}
\eeq
Therefore, Equation \eqref{SWB7} becomes
a {transcendental} equation for $\aKQ_2$ that needs to be solved
numerically.

{In summary, Equations \eqref{c5}--\eqref{c7bis}, \eqref{SWB7} and \eqref{Q0} provide the coefficients  $\SE_1$, $\SE_2$,
$\SE_3$, $\aQ$, $\aKQ_1$ and $\aKQ_2$ as functions of $\eta$, $T^*$ and $\lambda$. This in turn determines the Laplace transform $G(s)$ via Equations \eqref{2.2} and \eqref{eq:F(t)}. The rdf $g(r)$ can be obtained by numerically inverting $G(s)$ or, alternatively, by means of Equations \eqref{g(r)} and \eqref{varphi}.}

As a further asset of our present formulation, we must point out that, once $G(s)$ is determined, the static structure factor $S(q)$ (where $\mathbf{q}$ is the wavevector ) of the SS fluid may be readily obtained as
\beqa
\label{S(q)}
S(q)&=&1+\rho \int d\mathbf{r} \,e^{-\text{i} \mathbf{q}.\mathbf{r}}[g(r)-1]\nn
&=& 1- 2 \pi \rho \left.\frac{G(s)-G(-s)}{s}\right|_{s=\text{i}q},
\eeqa
where $\text{i}$ is the imaginary unit.

Finally, as proven in the Appendix, the RFA proposal \eqref{eq:F(t)} reduces to the exact solution of the PY equation \cite{W63,B68} in {three independent HS
limits: (a) $\lambda\to 1$,  (b) $T^*\to \infty$ and (c) $T^*\to 0$}.

{Before closing this section a further comment is in order. Our RFA approach has been based upon the exact rdf, $g_{\text{ex}}(r)$,  at zero density [cf.\ Equations \eqref{SSb}--\eqref{SW1}]. On the other hand, it can be shown that, to first order in density,  $g_\text{RFA}(r)$ differs  from $g_{\text{ex}}(r)$ in the region $1\leq r\leq\l$. More specifically \cite{YS94}},
\beqa
\lim_{\eta\to 0}\frac{\Delta g(r)}{\eta}&=&
e^{-1/T^*}\left(1-e^{-1/T^*}\right)(\l-1)\nn
&&\times\frac{\l-r}{r}\left[(\l-1)^2-3(\l+1)(r-1)\right]\nn
&&\times
\Theta(r-1)\Theta(\l-r),
\eeqa
{where $\Delta g(r)\equiv  g_\text{RFA}(r)-g_{\text{ex}}(r)$. This drawback could be corrected by a \emph{modified} RFA (mRFA) of the form \cite{MYS07}}
\beq
g_{\text{mRFA}}(r)=g_{\text{RFA}}(r)\exp\left[-\eta Q(r)\Theta(\l-r)\right]
\eeq
with
\beq
Q(r)=\frac{\l-r}{r}\left[Q_0-Q_1(r-1)\right],
\eeq
{such that $\lim_{\eta\to 0} Q_0=\left(1-e^{-1/T^*}\right)(\l-1)^3$ and $\lim_{\eta\to 0} Q_1=3\left(1-e^{-1/T^*}\right)(\l^2-1)$. The coefficients $Q_0$ and $Q_1$ for finite $\eta$ could be determined by imposing two extra conditions, such as the continuity of $y'(r)$ and $y''(r)$ at $r=\l$. Nevertheless, {{since such conditions again lead to technical complications and their usefulness is not known a priori,}} for the sake of simplicity we restrict ourselves {{here}} to the unmodified RFA described by Equations \eqref{2.2}, \eqref{eq:F(t)}--\eqref{c7bis}, \eqref{SWB7} and \eqref{Q0}. } {{The limitations imposed by this restriction may be assessed from the examination of the final results.}}

\section{Comparison with simulation data {and integral equation theories}\label{sec3}}

In order to assess the value of our theoretical approximation for the structural properties of SS fluids, in this section we perform a comparison between the results obtained with our approach and those obtained both from simulation and from our  {own} numerical solution of the PY {and HNC integral equations}.

As far as we know, the available simulation data for the rdf of the SS fluid as a function of distance at various packing fractions are those of
{Lang {\etal}} \cite{LKLLW99},
Zhou and Solana \cite{ZS09}
and
Guill\'en-Escamilla {\etal} \cite{GSC10}}. Although we have made a comparison with all these data, in Figs.\ \ref{fig1}--\ref{fig8} we only show graphs of $g(r)$ vs $r$ for some representative cases. Since our development is inspired by {and reduces to} the form of the solution of the PY equation for HS fluids {(see the Appendix)},  we have also included in the figures the results that we have obtained from the numerical solution of the PY equation for the SS fluid, {as well as those stemming out of the HNC equation for the same system.\footnote{The numerical solutions were obtained by solving the system of algebraic equations resulting from the discretisation of the integral equation. The convergence of the solutions  was found to be acceptable for a grid size of $\Delta r =0.0125$ and a cut-off distance $r_\text{max}=4$.}}

\begin{figure}
\begin{center}
\includegraphics[width=.8\columnwidth]{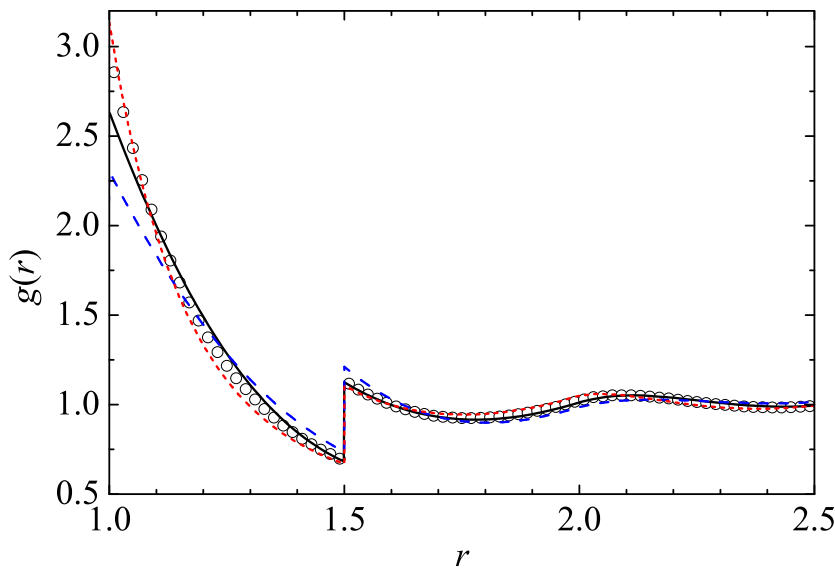}
\caption{  Radial distribution function $g(r)$ as a function of distance
$r$ for an SS fluid having $\lambda=1.5$, $T^*=2$ and  $\eta=0.3142$ ($\rho\sigma^3=0.6$) as obtained from the RFA approach (solid line), the PY equation (dashed line), the HNC equation (dotted line) and simulation data from  \protect\cite{GSC10} (circles).\label{fig1}}
\end{center}
\end{figure}

\begin{figure}\begin{center}
\includegraphics[width=.8\columnwidth]{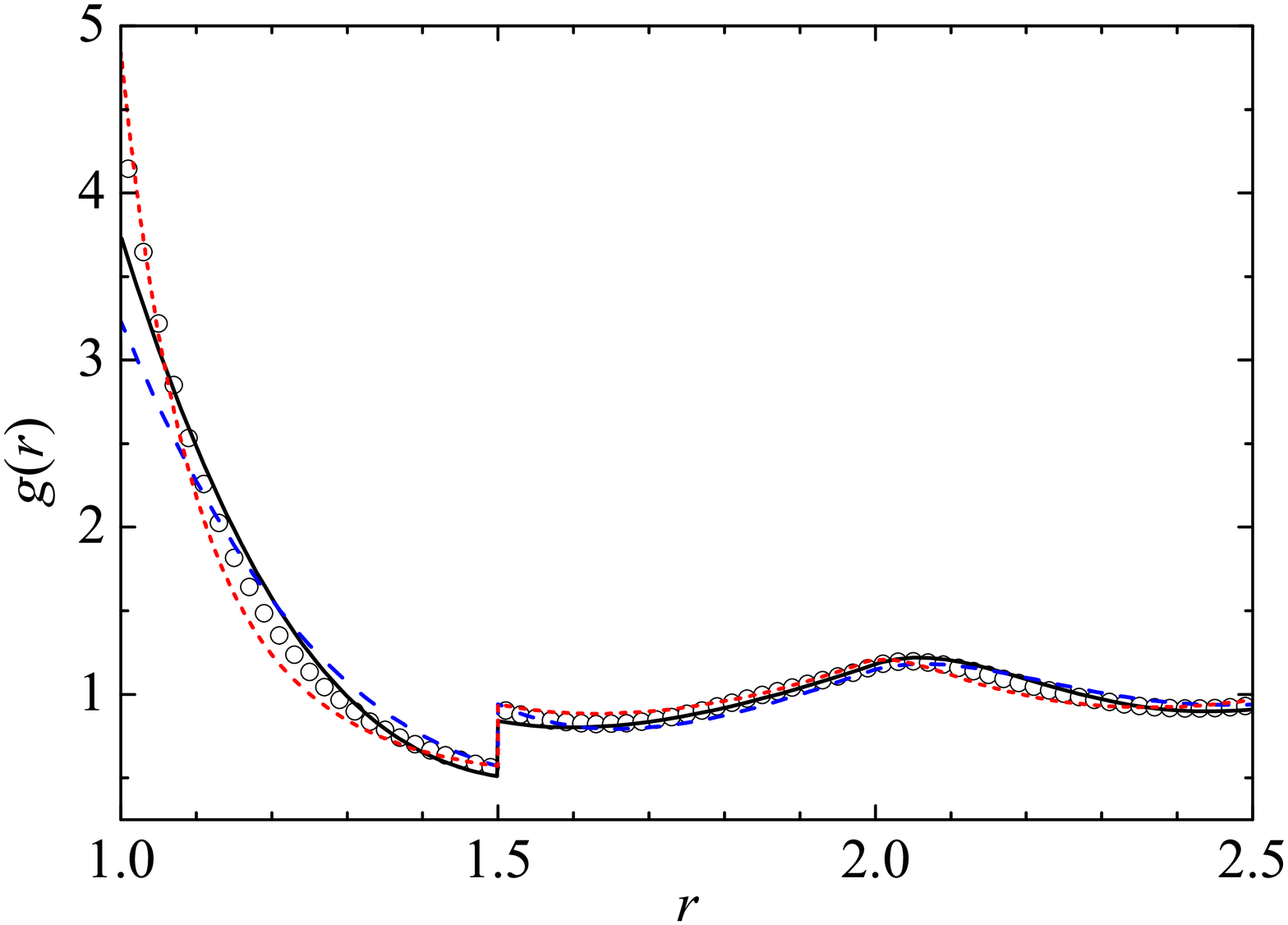}
\caption{  Radial distribution function $g(r)$ as a function of distance
$r$ for an SS fluid having $\lambda=1.5$, $T^*=2$ and $\eta=0.4189$ ($\rho\sigma^3=0.8$) as obtained from the RFA approach (solid line), the PY equation (dashed line), the HNC equation (dotted line) and simulation data from  \protect\cite{GSC10} (circles).}
\label{fig2}
\end{center}
\end{figure}

\begin{figure}\begin{center}
\includegraphics[width=.8\columnwidth]{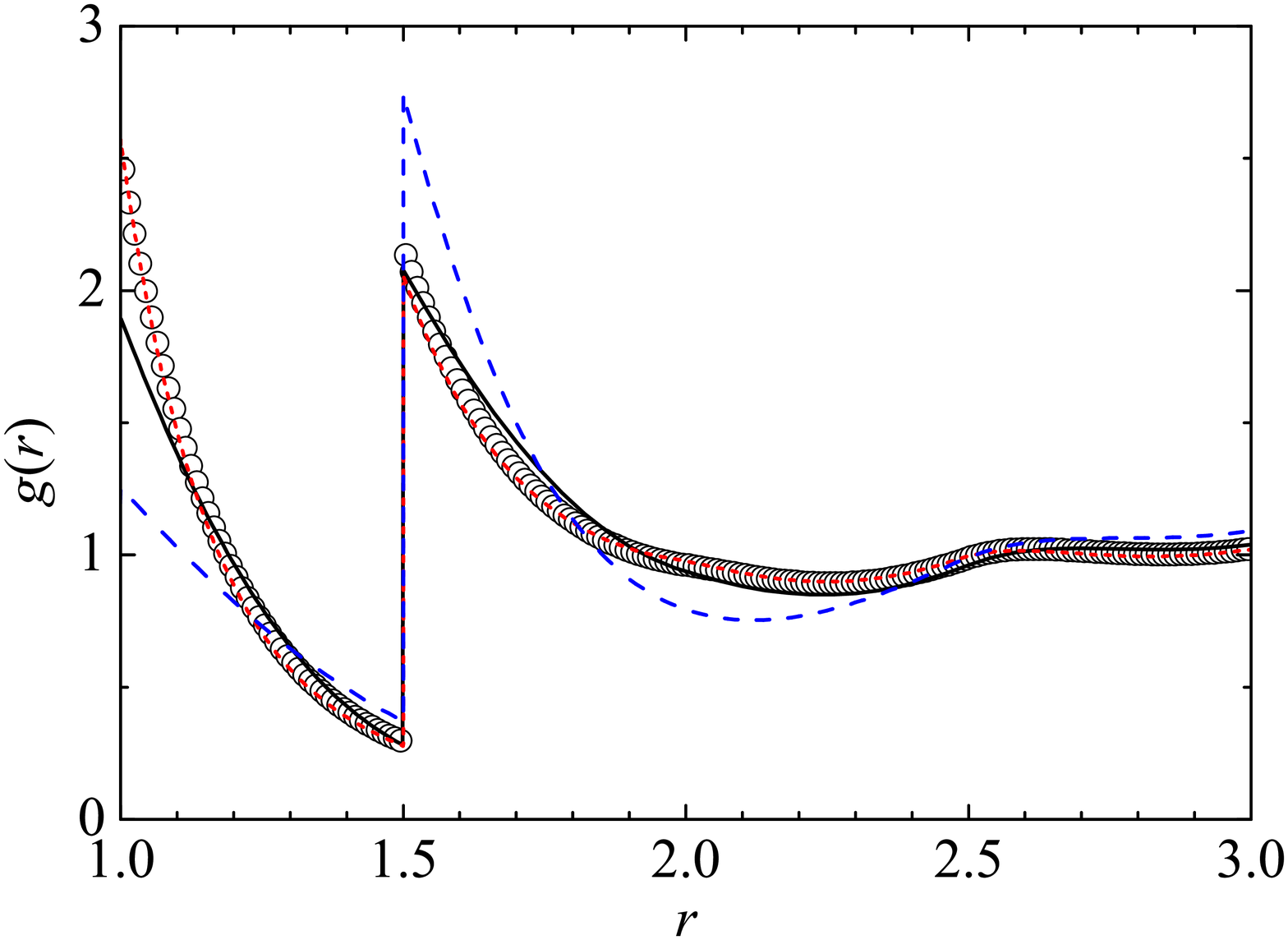}
\caption{  Radial distribution function $g(r)$ as a function of distance
$r$ for an SS fluid having $\lambda=1.5$, $T^*=0.5$ and $\eta=0.2094$ ($\rho\sigma^3=0.4$) as obtained from the RFA approach (solid line), the PY equation (dashed line), the HNC equation (dotted line) and simulation data from  \protect\cite{ZS09} (circles).}
\label{fig3}
\end{center}
\end{figure}

\begin{figure}\begin{center}
\includegraphics[width=.8\columnwidth]{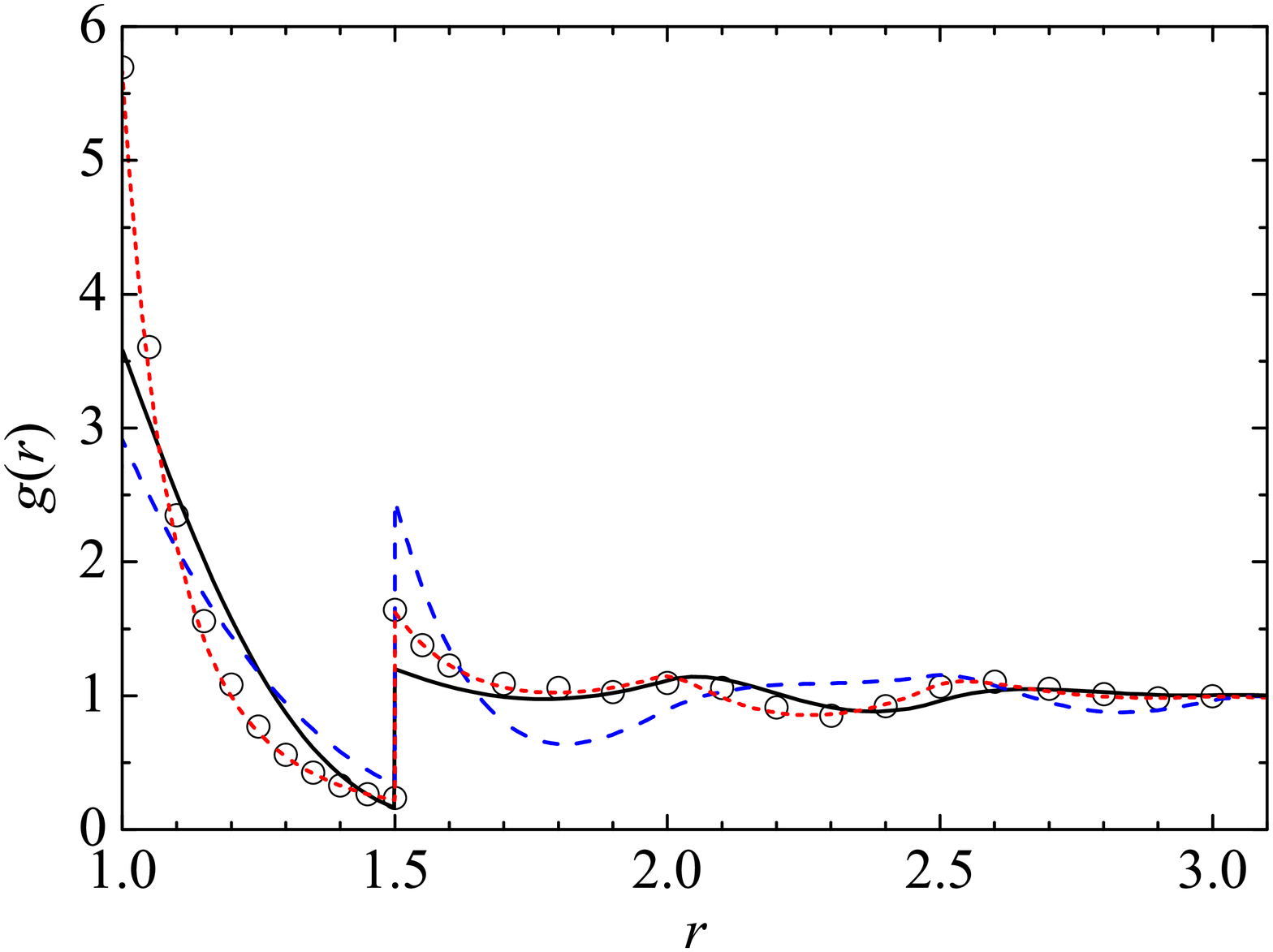}
\caption{  Radial distribution function $g(r)$ as a function of distance
$r$ for an SS fluid having $\lambda=1.5$, $T^*=0.5$ and $\eta=0.4$ as obtained from the RFA approach (solid line), the PY equation (dashed line), the HNC equation (dotted line) and simulation data from  \protect\cite{LKLLW99} (circles).}
\label{fig4}
\end{center}
\end{figure}

\begin{figure}\begin{center}
\includegraphics[width=.8\columnwidth]{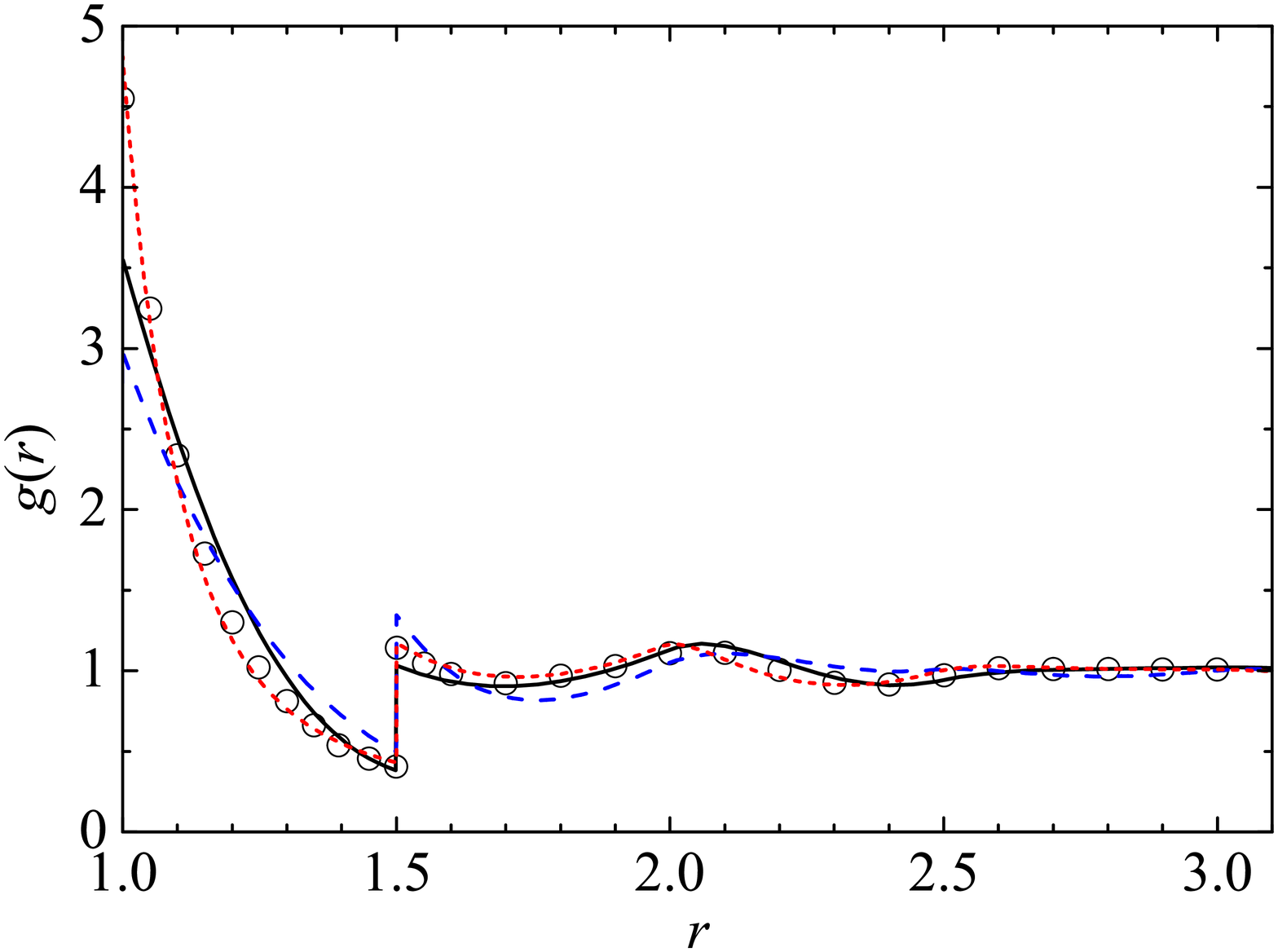}
\caption{  Radial distribution function $g(r)$ as a function of distance
$r$ for an SS fluid having $\lambda=1.5$, $T^*=1$ and $\eta=0.4$ as obtained from the RFA approach (solid line), the PY equation (dashed line), the HNC equation (dotted line) and simulation data from \protect\cite{LKLLW99} (circles).}
\label{fig5}
\end{center}\end{figure}

\begin{figure}\begin{center}
\includegraphics[width=.8\columnwidth]{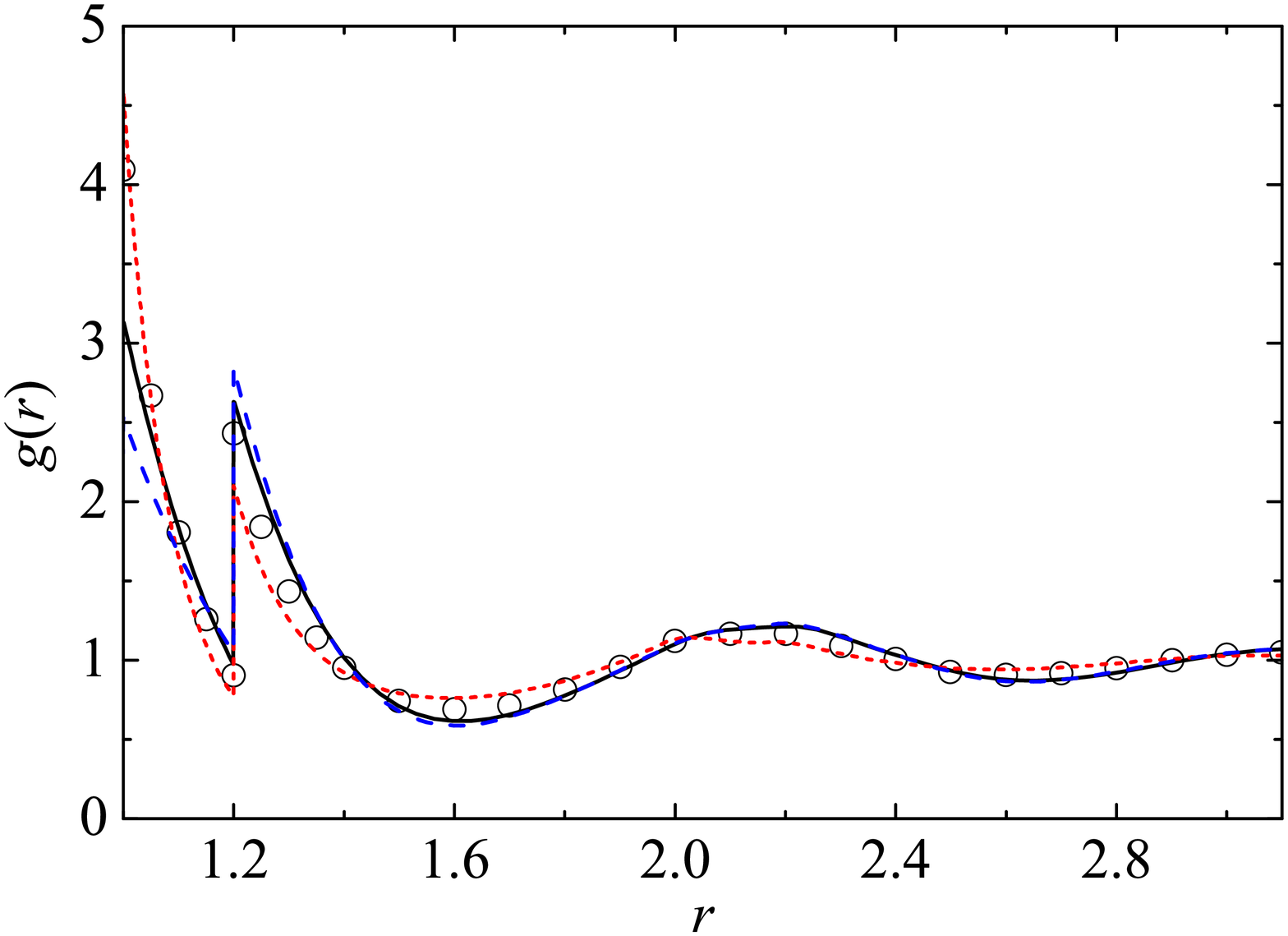}
\caption{  Radial distribution function $g(r)$ as a function of distance
$r$ for an SS fluid having $\lambda=1.2$, $T^*=1$, $\eta=0.4$ and  as obtained from the RFA approach (solid line), the PY equation (dashed line), the HNC equation (dotted line) and simulation data from  \protect\cite{LKLLW99} (circles).}
\label{fig6}
\end{center}\end{figure}

\begin{figure}\begin{center}
\includegraphics[width=.8\columnwidth]{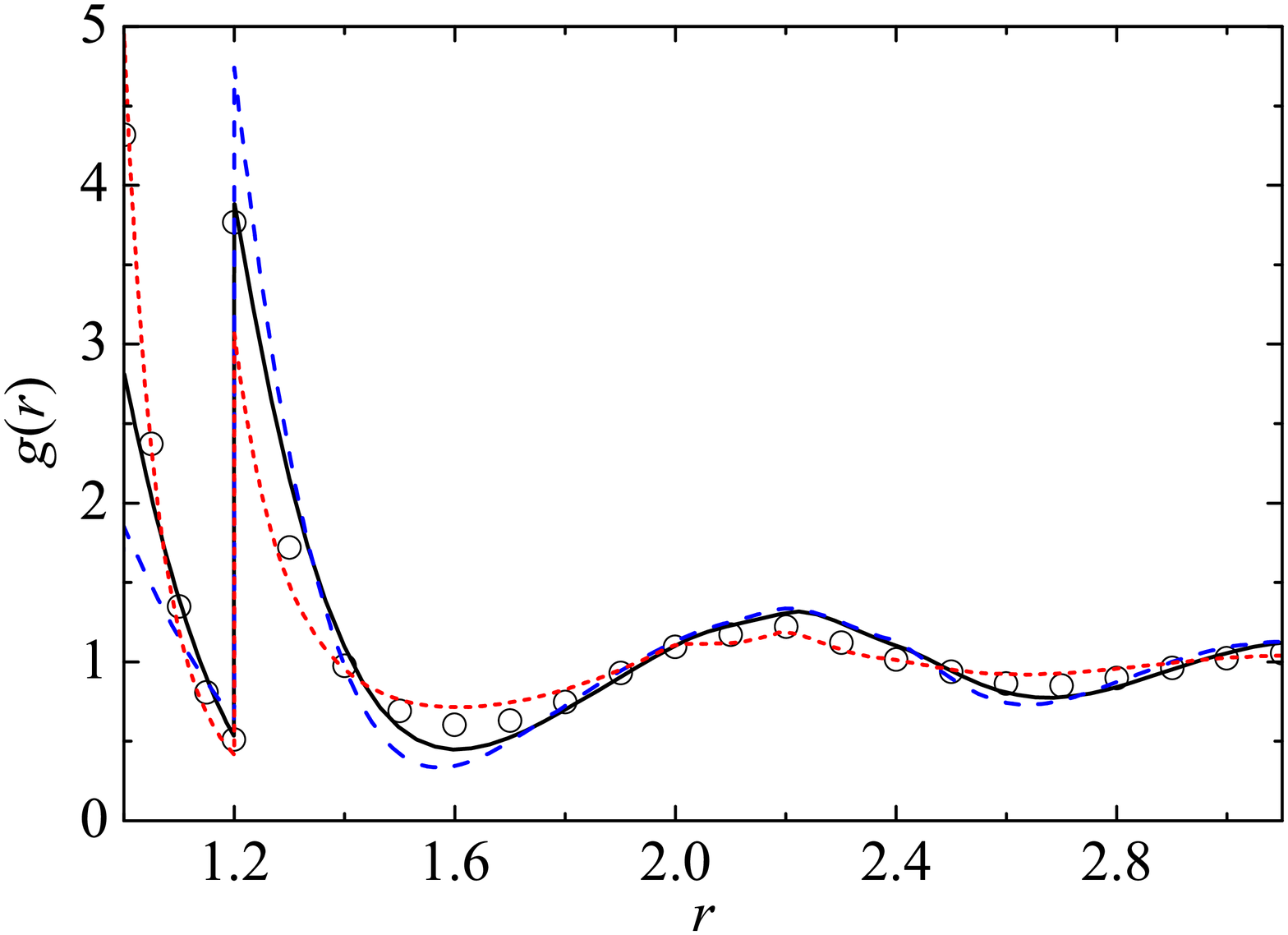}
\caption{  Radial distribution function $g(r)$ as a function of distance
$r$ for an SS fluid having $\lambda=1.2$, $T^*=0.5$ and $\eta=0.4$ as obtained from the RFA approach (solid line), the PY equation (dashed line), the HNC equation (dotted line) and simulation data from  \protect\cite{LKLLW99} (circles).}
\label{fig7}
\end{center}\end{figure}

\begin{figure}\begin{center}
\includegraphics[width=.8\columnwidth]{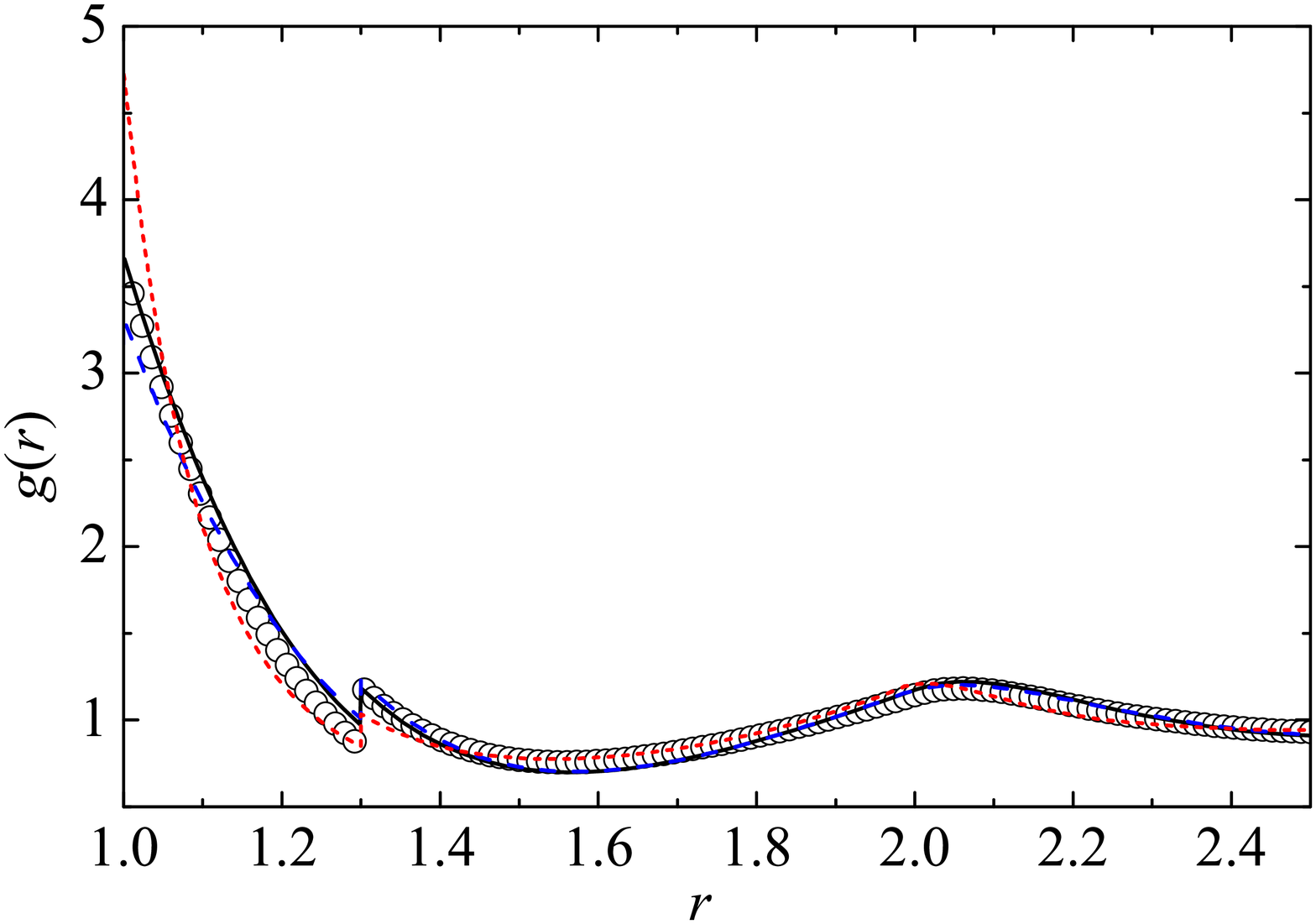}
\caption{  Radial distribution function $g(r)$ as a function of distance
$r$ for an SS fluid having $\lambda=1.3$, $T^*=5$ and $\eta=0.4189$ ($\rho\sigma^3=0.8$) as obtained from the RFA approach (solid line), the PY equation (dashed line), the HNC equation (dotted line) and simulation data from  \protect\cite{GSC10} (circles).}
\label{fig8}
\end{center}\end{figure}

{We start by fixing a shoulder width parameter $\l=1.5$ and  a relatively high reduced temperature $T^*=2$. Figures \ref{fig1} and \ref{fig2} show $g(r)$ for packing fractions $\eta=0.3142$ and $\eta=0.4189$, respectively. We observe a general good behaviour of the RFA, except that it underestimates the rdf near $r=1$. Interestingly, the performance of the RFA is clearly better than that of the numerical solution of the PY integral equation. On the other hand, the numerical solution of the HNC integral equation shows an excellent agreement with the simulation data.}

{The above features become much more apparent at a relatively low reduced temperature $T^*=0.5$, as Figs.\ \ref{fig3} and \ref{fig4} show. At $\eta=0.2094$ the limitations of the RFA are essentially restricted to the contact region $r\gtrsim 1$. However, the RFA becomes much less reliable at the larger density $\eta=0.4$. Again, the PY is rather poor, even at  $\eta=0.2094$, while the HNC keeps being very good.}

{It is interesting to analyse the influence of temperature for a fixed value of $\l$ and $\eta$. This can be achieved by comparing Figs.\ \ref{fig2}, \ref{fig4} and \ref{fig5} (even though the packing fraction in the case of Fig.\ \ref{fig2} is not exactly the same as that of Figs.\ \ref{fig4} and \ref{fig5}). As the temperature decreases, the contact value increases moderately and also $g(r)$  for $r\gtrsim 2$ becomes more structured, especially  when going from $T^*=1$ (Fig.\ \ref{fig5}) to $T^*=0.5$ (Fig.\ \ref{fig4}). The strongest influence of temperature occurs in the region  $r\approx\l$, the discontinuity  at $r=\l$ being much more pronounced as the temperature decreases, as expected on physical grounds.}

{To assess the influence of the shoulder width we consider the choice $\l=1.2$ in Figs.\ \ref{fig6} and \ref{fig7}, both at a common packing fraction $\eta=0.4$. As before, the RFA underestimates $g(r)$ near $r=1$, this effect being more important in the case of the PY theory. However,  the HNC solution is very reliable in that region. Upon comparison between Figs.\ \ref{fig5} and \ref{fig6}, on the one hand, and Figs.\ \ref{fig4} and \ref{fig7}, on the other hand, one may observe that shrinking the shoulder at fixed temperature and density makes the RFA and, to a lesser extent, the PY  approximation become more reliable, while the HNC approximation is slightly less accurate.}

{The above conclusion is consistent with the fact that the SS model becomes closer and closer to the HS model as the shoulder width decreases. In this HS limit the RFA reduces to Wertheim--Thiele's \cite{W63} exact solution of the PY equation (see the Appendix) and it is well known that such a solution is much more accurate than the HNC numerical solution for the HS fluid. Since the HS potential is also reached from  the the SS one in the high-temperature limit ($T^*\to\infty$), a better performance of  both RFA and PY over HNC  can be expected to hold for sufficiently high temperatures. This is confirmed by Fig.\ \ref{fig8} in the case $\l=1.3$, $T^*=5$ and $\eta=0.4189$ ($\rho\sigma^3=0.8$). In this state, the HNC theory clearly overestimates the contact value, a characteristic feature of the HS system. The SS model also becomes equivalent to the HS model (this time with a hard-core diameter $\l\sigma$) in the zero temperature limit ($T^*\to 0$). Therefore, again the RFA and PY predictions should be superior to the HNC ones for sufficiently low temperatures. {}From a practical point, however, this requires temperatures so low that $g(r)\approx 0$ for $1\leq r\leq \l$, what is obviously not the case of $T^*=0.5$, as Figs.\ \ref{fig3}, \ref{fig4} and \ref{fig7} show. }

{In summary,} from the analysis of the results it is clear that the RFA approach represents a clear improvement as compared to the PY approximation.  {It also does a fair job in comparison with the HNC equation, which is generally the best approximation. On the other hand,} this latter becomes more inaccurate, particularly at the contact value, as the HS limit is approached, in contrast to the good performance of the RFA approach in this limit. Also worth mentioning is the fact that, {although not shown, the RFA also beats the numerical solution of the OZ relation closed by the hard-core condition $g(r)=0$ for $r<1$ plus the parametrization  of the direct correlation function for $r>1$  proposed by Guill\'en-Escamilla {\etal}
\cite{GSC10}}. As expected, although {the RFA} always underestimates the contact value, the present approximation works particularly well at all distances for narrow shoulders at high temperatures and low densities. As the shoulder width increases, for a fixed (relatively high) temperature, the packing fraction at which deviations become more pronounced decreases. In the case of low temperatures, the trend observed for low densities is maintained but this time the region between contact and the shoulder width is described more poorly, while for greater distances the approach  seems to capture rather well the subsequent oscillations of the rdf.

\section{Concluding remarks}
\label{sec4}

In this paper we have presented an (almost completely) analytical procedure to obtain the structural properties of SS fluids. {Although the derivation heavily  relied upon a parallel development for the SW fluid \cite{YS94},}   the results could not be anticipated since the {SW and SS potentials are physically quite different: while the SW potential has an attractive part that allows for the existence of a vapour-liquid phase transition ending at a critical point, the SS potential is purely repulsive}.

Our procedure, which follows the same rationale that we have used for other systems \cite{HYS08}, is inspired on the analytical solution of the PY equation for HS fluids \cite{W63,B68} {(to which it reduces in the appropriate limits)} and represents a good alternative to the usual integral equation approach of liquid state theory, {which requires numerical work}. As a matter of fact, as shown above, it clearly provides an improvement over the results of the simplest such integral equation for SS fluids, namely the OZ equation with the PY closure, {and compares reasonably well with the results of the HNC integral equation, which in particular it beats {near} the HS limits}.

The importance of analytical or semi-analytical approximations for the
equilibrium structural properties of simple fluids cannot be overemphasized. In
this respect, we find it especially remarkable the fact that our approach, which
only requires the solution of a single transcendental equation when the SS
potential differs from the HS one (i.e., if $\lambda\neq 1$ and $T^*$ is finite), behaves
better than the PY integral equation, whose solution in the same situation,
however, involves much more numerical work.

Through the comparison with the available simulation data \cite{LKLLW99,ZS09,GSC10}, we have been able to identify roughly the region in the $(\eta,T^*,\l)$ space that these data span where the approximation  for $g(r)$ yields a reasonably good performance. The RFA  approach for SS fluids leads to rather accurate results at any fluid
density if the shoulder is sufficiently narrow (say $\l \leq 1.2$), as well as for any width if the density is small enough ($\eta \leq 0.4$). However, as the width and/or the density increase, the RFA predictions  worsen, especially at low temperatures and in the region between contact and $\l$. {{In any event, it is fair to insist on the advantage of the RFA quasi-analyticity and its relative good numerical results at low and high fluid densities for the SS fluid.}}

Apart from the rdf, our proposal allows the immediate computation of the static structure factor, {as shown by Equation \eqref{S(q)}}. We are not aware of the existence of simulation data for this structural property {in the case of the SS interaction} so a comparison is not feasible at this stage. We hope that our work may serve as a motivation to carry out such simulations. Also worth pointing out is the fact that the availability of the rdf allows us to obtain an approximate {direct correlation function $c(r)$  through inversion of the OZ equation in Fourier space}. {{Once more, we are not aware of any simulation results for the $c(r)$ of this system one could compare with. Given this situation and since these structural properties do not exhibit any particular feature we have not presented plots of them. However they may be easily produced upon request.

The present results suggest that the study of the  structural properties of fluids whose particles interact via discrete potentials composed of combinations of square wells and square shoulders, and other piece-wise potentials, may be tackled in a similar way. We plan to carry out some of these studies in future work. Finally, we also plan to examine in the near future the prediction and discussion of the thermodynamics of SS fluids which follows from our approach.}}

\section*{Acknowledgements}

We want to thank C. N. Likos, J. R. Solana and R. Casta\~{n}eda-Priego for sending us their simulation data. Two of us (S.B.Y. and A.S.) acknowledge the financial support of the Ministerio de Ciencia e Innovaci\'on (Spain) through Grant No. FIS2010-16587 (partially financed by FEDER funds) and of the Junta de Extremadura through Grant No. {GR10158}. The work of M.L.H. has been partially supported by DGAPA-UNAM under project IN-107010-2. He also wants to thank the hospitality of Universidad de Extremadura, where the first {stages of this work were carried out}.

\appendices
\section{The hard-sphere limit}
\label{appA}
In this appendix we will show that the proposal we have introduced for the structure of the SS fluid reduces, in the appropriate limits, to the PY approximation for HS fluids. We begin with the latter. For an HS fluid, the Laplace transform $G^\HS(s;\e)$ of $rg^\HS(r)$ in the PY approximation, where we have made it explicit that it depends on both $s$ and the packing fraction $\e$, may be expressed as in Equation (\ref{2.2}), where the auxiliary function $\Psi^\HS(s;\e)$ takes { a form similar to that  of Equation (\ref{eq:F(t)}), namely} \cite{YS91,W63}
\begin{equation}
\Psi^\HS (s;\e)=\frac{1}{2\pi}\frac{-12\eta+\SE_1^\HS(\e) s+\SE_2^\HS(\e) s^2+\SE_3^\HS(\e)
s^3}{1+\overline{\aKQ}_1^\HS(\e) s},
\label{PsiHS}
\end{equation}
with
\beq
\overline{\aKQ}_1^\HS(\e)=\frac{1+{\e}/{2}}{1+ 2 \e},
\label{Q1HS}
\eeq
\beq
\SE_1^\HS(\e)=\frac{18\e^{2}}{1+ 2 \e},
\label{E1HS}
\eeq
\beq
\SE_2^\HS(\e)=\frac{6 \e \left(1-\e\right)}{1+ 2 \e},
\label{E2HS}
\eeq
 and
\beq
\SE_3^\HS(\e)=\frac{\left(1-\e\right)^2}{1+ 2 \e}.
\label{E3HS}
\eeq

For the sake of clarity, it is convenient in the case of the SS fluid to include explicitly the dependence on the packing fraction $\eta$, the {interaction range} $\l$, and the temperature parameter $T^*$ in the function  $G(s)$  through the notation $G(s;\e,\l,T^*)$.

\subsection{Limit $\l\to 1$}
We must clearly
recover the HS case if $\l=1$, in which case the SS potential becomes equivalent to an HS
interaction of diameter $\sigma=1$, {i.e.}
\beq
G(s;\e,1,T^*)=G^\HS(s;\e).
\label{N1}
\eeq

{It is straightforward to see from Equations \eqref{c6}--\eqref{c7bis} that in the limit $\l\to 1$  one has $\SE_i(\e,\l,T^*)\to \SE_i^\HS(\e)$, irrespective of the values of $\lim_{\l\to 1} \aQ(\e,\l,T^*)$ and $\lim_{\l\to 1} \aKQ_2(\e,\l,T^*)$. On the other hand, the denominator of Equation \eqref{eq:F(t)} becomes $1+\left[\aKQ_1(\e,1,T^*)+\aKQ_2(\e,1,T^*)\right]s=1+\overline{\aKQ}_1^\HS(\e) s$ on account of Equation \eqref{c5}. This completes the proof that Equation \eqref{eq:F(t)} reduces to Equation \eqref{PsiHS} in the limit $\l\to 1$ for arbitrary $T^*$.}

\subsection{Limit $T^*\to \infty$}

{The high-temperature limit $T^*\to\infty$ can be understood as the limit $\epsilon\to 0$ at finite $T$, so the SS potential \eqref{SS} trivially becomes that of HS. Consequently, }
\beq
G(s;\e,\l,\infty)=G^\HS(s;\e).
\label{N1b}
\eeq
{In the limit $T^*\to\infty$ one has $e^{1/T^*}\to 1$, so that Equations \eqref{SWB7} and \eqref{Q0} imply $\aQ(\e,\l,\infty)=\aKQ_2(\e,\l,\infty)=0$. Next, Equations \eqref{c5}--\eqref{c7bis} yield $\aKQ_1(\e,\l,\infty)=\overline{\aKQ}_1^\HS(\e)$ and $\SE_i(\e,\l,\infty)= \SE_i^\HS(\e)$ for any $\l$.}

\subsection{Limit $T^*\to 0$}

{On the other hand, the low-temperature limit $T^*\to 0$ corresponds to the limit $\epsilon\to \infty$ at finite $T$. In that case,} the SS potential becomes equivalent to an HS interaction of diameter $\l\sigma=\l$. This latter condition  implies that
\beq
G(s;\e,\l,0)=\l^2 G^\HS(\l s;\l^3\e),
\label{N2}
\eeq
which is a non-trivial scaling relation. In turn, {taking into account the relationship \eqref{2.2},} Equation (\ref{N2}) leads to
\beq
\Psi(s;\e,\l,0)= \frac{e^{(\l-1)s}}{\l^3}\Psi^\HS(\l s;\l^3\e).
\label{N3}
\eeq

{}From Equations \eqref{eq:F(t)} and (\ref{PsiHS}) it follows that Equation \eqref{N3} is satisfied if the coefficients  comply with the following conditions:
\beq
\aQ(\e,\l,0)=1,
\label{N4}
\eeq
\beq
\aKQ_1(\e,\l,0)=0,
\label{N6}
\eeq
\beq
\aKQ_2(\e,\l,0)=\l \overline{\aKQ}_1^\HS(\l^3\e),
\label{N5}
\eeq
\beq
\SE_1(\e,\l,0)=\l^{-2} \SE_1^\HS(\l^3\e),
\label{N7}
\eeq
\beq
\SE_2(\e,\l,0)=\l^{-1} \SE_2^\HS(\l^3\e),
\label{N8}
\eeq
\beq
\SE_3(\e,\l,0)=\SE_3^\HS(\l^3\e).
\label{N9}
\eeq

{Equation \eqref{N4} is a direct consequence of Equation \eqref{Q0}. Next, Equation \eqref{SWB7} in the limit $T^*\to 0$ yields Equation \eqref{N6}. Then, by imposing $\aKQ_1=0$ in Equation \eqref{c5} one gets Equation \eqref{N5}. Finally, by inserting Equations \eqref{N4} and \eqref{N5} into Equations \eqref{c6}--\eqref{c7bis} it is easy to check that Equations \eqref{N7}--\eqref{N9} are  verified.}

\end{document}